\documentclass[preprint,superscriptaddress,showkeys]{revtex4-1}
\usepackage{geometry}
\usepackage{xcolor}
\usepackage{tabu}
\usepackage{titlesec}
\usepackage[utf8]{inputenc}
\usepackage{graphicx,epsfig,epstopdf}
\usepackage{bm}
\usepackage{epstopdf}
\usepackage{amssymb,amsmath,amsfonts,latexsym}
\usepackage{dcolumn}
\usepackage{bm}
\usepackage{hyperref}
\usepackage[utf8]{inputenc}

\usepackage{breqn}             
\makeatletter                  
\let\cat@comma@active\@empty   
\makeatother                   

\begin{document}
\title{Bottomonium spectroscopy motivated by general features of pNRQCD}%

\author{Raghav Chaturvedi}
\email{raghavr.chaturvedi@gmail.com}
\affiliation{Department of Applied Physics, Sardar Vallabhbhai National
Institute of Technology, Surat$-$395007, Gujarat, \textit{INDIA}.}

\author{A. K. Rai}
\email{raiajayk@gmail.com}
\affiliation{Department of Applied Physics, Sardar Vallabhbhai National
Institute of Technology, Surat$-$395007, Gujarat, \textit{INDIA}.}

\author{N. R. Soni}
\email{nrsoni-apphy@msubaroda.ac.in}
\affiliation{Physics Department, Faculty of Science, The Maharaja Sayajirao University of Baroda, Vadodara 390002, Gujarat, {\it INDIA}.}

\author{J. N. Pandya}
\email{jnpandya-apphy@msubaroda.ac.in}
\affiliation{Applied Physics Department, Faculty of Technology and Engineering, \\ The Maharaja Sayajirao University of Baroda, Vadodara 390001, Gujarat, {\it INDIA}.}

\date{\today}
\begin{abstract}
The bottomonium mass spectra is computed in the framework of potential non-relativistic quantum chromodynamics. The potential consists of a static term incorporating Coulombic plus confinement part along with a correction term added non-perturbatively from pNRQCD, which is classified in powers of the inverse of heavy quark mass \textit{O}$(1/m)$. The masses of excited bottomonia are calculated by perturbatively adding spin-hyperfine, spin-orbit and tensor components of one gluon exchange interactions in powers of \textit{O}$(1/m^2)$. Calculated masses are found to be consistent with other theoretical studies and experimental data. The Regge trajectories of the calculated mass spectra are also constructed. The values of the wave-functions are extracted and employed to calculate the electromagnetic transition widths and $\gamma\gamma$, $e^+e^-$, light hadron and $\gamma\gamma\gamma$ decay widths of several states at various leading orders, within the non-relativistic QCD formalism.
Some of the experimentally reported states of bottomonium family like $\Upsilon$(10860), $\Upsilon$(11020) and X(10610) are identified as mixed $S-D$ wave and $P$-wave states.\end{abstract}
\vspace{2pc}

\maketitle

\section{Introduction}
Bottomonium states $\Upsilon(1S-3S)$ were discovered at Fermilab in $1977$ \cite{Herb:1977ek,Innes:1977ae}. Three decades later in 2008, after many experimental attempts, spin singlet state $\eta_b(1S)$ was successfully identified by Belle collaboration \cite{Mizuk:2012pb} followed by observation of $\eta_b(2S)$ by BABAR \cite{Lees:2011mx}, CLEO \cite{Dobbs:2012zn} and Belle \cite{Sandilya:2013rhy}. The lowest spin singlet $P$-state $1^1P_1$ was also observed by BABAR \cite{Lees:2011zp}. The proton anti-proton colliders have also joined along with Belle, BABAR, CLEO, ATLAS \cite{Aad:2011ih} and are expected to produce more precise data. Bottomonium system is believed to have non-relativistic nature and hence can be treated as two body system of heavy quark and antiquark. At present, there are $21$ experimentally known states of bottomonia, therefore the investigation of the masses of bottomonium states from theoretical point of view is essential. These studies will allow to single out experimental candidates and would prove to be powerful tools for understanding the quark anti-quark interaction as expected from quantum chromodynamics (QCD). Comprehensive reviews on the status of theoretical as well as the experimental status of heavy quarkonium physics are found in literature \cite{Eichten:2007qx,Godfrey:2008nc,Barnes:2009,Brambilla:2010cs,Brambilla:2014jmp,Andronic:2015wma}.
Certain approaches like lattice QCD \cite{McNeile:2012qf}, chiral perturbation theory \cite{Gasser:1983yg}, heavy quark effective field theory \cite{Neubert:1993mb}, non-relativistic effective field theory \cite{Brambilla:2019jfi}, QCD sum rules \cite{Veliev:2010vd}, NRQCD \cite{Mateu:2018zym}, dynamical equations based approaches like Schwinger-Dyson and Bethe-Salpeter equations (BSE) \cite{Chen:2020ecu,Mitra:2001ns,Alkofer:2000wg,Ricken:2000kf,Mitra:1990av},
relativistic flux tube model \cite{Chen:2019uzm},
an effective super-symmetric approach \cite{Nielsen:2018uyn}, quark models \cite{Li:2019tbn,Ortega:2020uvc} and various potential models have tried to explain the phenomena of quark confinement and dynamics of QCD.

Considering the complex structure of QCD vacuum, it is difficult to obtain quark-antiquark interaction potential starting from basic principles of QCD making it necessary to account for non-relativistic effects. Relativistic effects are not very prominent in case of the bottomonia states, also the binding energy is significantly small in comparison with the rest mass energy of the constituents. The theoretical uncertainty in the potential at large and intermediate distances had led to the beginning of potential models \cite{Eichten:2019gig,Ebert:2011jc,KumarRai:2005yr,Rai:2006bt,Rai:2008sc,Rai:2008zza,Parmar:2010ii,Rai:2017fui,Kher:2018wtv,Rathaud:2016tys,Vinodkumar:1999da,Pandya:2001hx}. The widely accepted practice to obtain the masses involves choice of non-relativistic Hamiltonian with a potential and solving the Schr\"{o}dinger equation numerically \cite{Lucha:1998xc}. Most common of all potential is the Coulomb plus linear confinement and it has been supported by static potential computed using lattice QCD simulations that offer powerful tools for the non-perturbative study of QCD. But,
along with calculation of mass spectra of any mesonic state, certain decay channels need to be tested as well to validate the model. Thus, the test for any theoretical model is to reproduce the mass spectra as well as decay properties. Certain relativistic and non-relativistic potential models have successfully computed mass spectra but were not consistent in prediction of the decay properties \cite{Quigg:1977dd,Martin:1980jx,Buchmuller:1980su,Kiselev:1994rc,Rai:2008sc,Parmar:2010ii}.
This inconsistency inspires us to calculate mass spectra of $b\bar{b}$ meson along with its decay properties. In the present study, in addition to the Coulomb plus confinement terms, a spin dependent relativistic correction term has been considered. This relativistic correction to the mass has been derived in \cite{Pineda:2000sz} in the framework of pNRQCD \cite{Pineda:1997bj,Brambilla:1999xf,Brambilla:2004jw,Brambilla:2020xod} which is an effective field theory, and is usually classified in powers of the inverse of heavy quark mass or velocity. In heavy quarkonia system, there exists a hierarchy $m\gg mv \gg mv^2$, with $m \gg \Lambda_{QCD}$ where $\Lambda_{QCD}$ is a QCD scale parameter assumed to be of the order of few hundreds of MeV. To have better control of such hierarchy, one needs to employ effective field theory (EFT) for high precision calculations. Two such EFT's have been used by many authors in past namely non-relativistic QCD (NRQCD) \cite{Caswell:1985ui,Brambilla:2010cs} and potential NRQCD (pNRQCD) \cite{Brambilla:2004jw}. NRQCD was derived by integrating the energy scale above $m$ in QCD, and pNRQCD by integrating further the energy scale above $mv$ in NRQCD. Higher energy contribution is incorporated in effective couplings called the matching coefficients.

Moreover, results in recent years have raised interest in heavy flavor physics \cite{Brambilla:2010cs,Tanabashi:2018oca,Soni:2017wvy} both in the region above and below $B\bar{B}$ threshold to study the strong interaction between hadrons. In the region above $B\bar{B}$ threshold, the masses beyond $4S$ are not well resolved, like so called $X$, $Y$ and $Z$ states that have unusual properties and might turn out to be exotic states, mesonic molecules, multi quark or even hybrid states \cite{Brambilla:2010cs}. In this article, we have also tried to emphasize on such states like $X(10610)$, $\Upsilon$(10860) and $\Upsilon$(11020) which share same $J^{PC}$ values as some of the states of bottomonium and justify to be one of them. Decay properties of $X(10610)$, $\Upsilon(10860)$ and $\Upsilon(11020)$ can also throw light on their identity.

This paper is organized as follows: After introduction, the theoretical framework is given in Section \ref{sec:Theoretical framework}. In Section \ref{sec:Regge}, details of Regge trajectories are given. Section \ref{sec:Decay} deals with $\gamma\gamma$, $e^{+}e^{-}$, light hadron and $\gamma\gamma\gamma$  decay widths of different states. In Section \ref{sec:Mixing}, details of some positive and negative parity states of bottomonium as a mixture $P$-wave and mixed $S-D$ wave states are discussed. In Section \ref{sec:EM}, details of Electromagnetic transition widths are discussed and finally in Sections \ref{sec:Result} and \ref{sec:Conclusion}, results, discussion and conclusion are presented.
\begin{table*}[!ht]
\caption{Experimental status of some negative and positive parity $b\bar b$ mesons near open-flavor threshold reported by PDG\cite{Tanabashi:2018oca}}.
\scalebox{1}{
\begin{tabular}{cccccc}
\hline\noalign{\smallskip}
\hline
PDG& Former/Common  & Expt.Mass  & J$^P$ & Production & Discovery\\
Name&Name&(in keV)&&&Year\\
\hline
$X(10610)$&${Z_{{b0}}{(10610)}}$&10609 $\pm$6 &$1^+$&${{\mathit \Upsilon}{(5S)}}\rightarrow{{\mathit \Upsilon}{(nS)}}{{\mathit \pi}^{+}}{{\mathit \pi}^{-}}$&2012\\
\hline
${{\mathit \Upsilon}{(10860)}}$&--&$10889.9^{+3.2}_{-2.6}$&$1^-$&${{\mathit e}^{+}}{{\mathit e}^{-}}\rightarrow hadrons$&1985\\
\hline
${{\mathit \Upsilon}{(11020)}}$&--&$10992.9 {}^{+10.0}_{-3.1}$&$1^-$&${{\mathit e}^{+}}{{\mathit e}^{-}}\rightarrow hadrons$&1985\\
\noalign{\smallskip}\hline\hline
\end{tabular}}
\label{Table:bbmix}
\end{table*}

\section{Theoretical framework}
\label{sec:Theoretical framework}
We consider bottomonia as non-relativistic system and in order to calculate its mass spectra, the following Hamiltonian is used in present study \cite{Gupta:1994mw}
  \begin{equation}
  H=M+\frac{P^2}{2 \mu}+ V_{pNRQCD}(r)+ V_{SD}(r)
  \end{equation}
Here, $M$ is the total mass of the system and $\mu$ is the reduced mass of the system.
Interaction potential $V_{pNRQCD}(r)$ encompasses three terms: a Coulombic term $V_v$(r) (vector), a confinement term $V_s$ (scalar) and relativistic correction $V_p(r)$ in the framework of pNRQCD  \cite{article}.
\begin{eqnarray}
\label{eq:potential}
V_{pNRQCD}(r) & = & V_v(r) + V_s(r) + V_p(r) \nonumber \\
 &=& -\frac{4\alpha_{s}}{3r}+ \frac{A}{n^{0.16}} r + \frac{1}{m_b}V^{(1)}(r)\\
V^{(1)}(r)&=& -\frac{9\alpha_{c}^2}{8r^2}+ a \log r + C
\end{eqnarray}
$\alpha_s$ and $\alpha_c$ are strong running coupling constants and effective running coupling constant respectively, $m_{b}$ is mass of bottom quark, $a$ and $C$ are potential parameters. $\alpha_s$ can be computed as
\begin{eqnarray}\label{eq:running_coupling}
    \alpha_s (\mu^2) = \frac{4 \pi}{(11-\frac{2}{3} n_f) \ln (\mu^2/\Lambda^2)}
\end{eqnarray}
where $n_f$ is the number of flavors, $\mu$ is renormalization scale related to the constituent quark masses as $\mu = 2 m_Q m_{\bar Q}/(m_Q + m_{\bar Q})$ and $\Lambda$ is a QCD scale which is taken as 0.15 GeV by fixing $\alpha_s$ = 0.1185 \cite{pdg2016} at the $Z$-boson mass. In order to obtain mass difference between degenerate mesonic states, we consider spin-dependent part of the usual one gluon exchange potential given by,
\begin{eqnarray}
V_{SD}(r) & = &\frac{1}{m_b^2} \bigg[V_{SS}(r)+ V_{L\cdot S}(r)+
V_T(r) \bigg [ S(S+1)-3(S\cdot \hat{r})(S\cdot \hat{r}) \bigg]\bigg]
\end{eqnarray}
Where the spin-spin, spin-orbital and tensor interactions are given as\cite{article}, with $C_f=\frac{4}{3}$.
\begin{eqnarray}
V_{SS}(r) = \frac{8}{9 }\frac{\alpha_s }{m_Q m_{\bar{Q}}}\overrightarrow{S}_Q \overrightarrow{S}_{\bar{Q}} 4 \pi \delta^3(\vec r),
\end{eqnarray}

\begin{eqnarray}
V_{L\cdot S}(r) &=& \frac{C_s}{2 r} \frac{d}{dr} (V_v(r) + V_s(r)) + \frac{C_f}{r}\left[-\left(1-\epsilon\right)\sigma +\left(\frac{\alpha_c}{r^2}+ \epsilon \sigma\right)\right]
\end{eqnarray}

\begin{eqnarray}
V_{T}(r) &=& \frac{{C_f}^2}{3} \frac{3 \alpha }{r^3}
\end{eqnarray}
We fix the mass of bottom quark and other potential parameters like $\alpha_c$, $\alpha_s$, $\epsilon$, $A$, $\alpha$, $\sigma$, $C$ and $a$, their values are given in Table \ref{Table:parameters}.

\begin{table}[!ht]
{\tiny \caption{Potential parameters}
\renewcommand{\arraystretch}{0.7}
\begin{tabular*}{\textwidth}{@{\extracolsep{\fill}}cccccccccc@{}}
\hline
Parameters&$\alpha_c$ & $\alpha_s$ & $m_b$ & $\epsilon$ & A & $\alpha$ & $\sigma$ & a & C\\
\hline
Present &0.19& 0.253& 4.81GeV & 0.2 & 0.28 $\frac{GeV}{fm}$ & 0.216 & 0.1 & 0.15 ${GeV}^2$ &0.523\\
work &&&&  & & &  & & \\
\hline
\cite{article}& 0.2227& 0.297& $4.18 {}^{+0.03}_{-0.02}$GeV \cite{Tanabashi:2018oca}  & 0.2 & 1.06 $\frac{GeV}{fm}$ & 0.216 & 0.2 & 0.142 ${GeV}^2$ &--\\
\hline
\end{tabular*}
\label{Table:parameters}}
\end{table}

Here, we first compute ground state mass of bottomonia state to fix quark mass and confinement strength after fitting the spin averaged ground state masses with experimental data of respective mesons. Values of the potential parameters are determined and used in computation of mass spectra. The Schr\"{o}dinger equation is solved numerically for pNRQCD potential Eq. (\ref{eq:potential}) by Runge-Kutta method using the \textit{Mathematica} notebook \cite{Lucha:1998xc}. In principle, the integration starts at origin ($r=0$) but because of singularity reason the integration has to start at some finite value close to but different from origin and sticking to the limitations of the boundary conditions. We have chosen closest nonzero distance between the quark and antiquark to be of the order of quark size as suggested in \cite{Shastry:2018oix}. Also, normalised wave function for respective states are determined numerically. The parameter $A$ represents potential strength analogous to spring tension. For the excited states of the bottomonia, we choose the size parameter to be state dependent as $\frac{A}{n^{0.16}}$. It can be justified by the similar arguments for the changes in $\alpha_s$ with the average kinetic energy \cite{Patel:2008na}. Here, as the system gets excited, the average kinetic energy increases and hence the potential strength (string tension) reduces. With this mild state dependence on the potential parameter $A$, we obtain the excited spectra as well as the right behavior for the radial wave-functions.  This scaling of potential strength shifts the radially excited state masses as well as spin-orbit splittings closer to experimental data \cite{Tanabashi:2018oca}.

Due to spin-dependent hyperfine interaction the radial wave-function for pseudoscalar and vector states will be different. This correction is included in the value of $R(0)$ by considering \begin{eqnarray}\label{eq:rc}
R_{nJ}(0)=R(0)\left[1+(SF)_J
\frac{\langle\varepsilon_{SD}\rangle_{nJ}}{M_{SA}} \right]\end{eqnarray} where $(SF)_J$
and $\langle\varepsilon_{SD}\rangle_{nJ}$ is the spin factor and spin
interaction energy of the meson in the $nJ$ state, while $R(0)$
and $M_{SA}$ correspond to the radial wave function at
the zero separation and spin average mass of the $Q \bar Q$ system respectively. It can be seen that Equation (\ref{eq:rc}) provides the radial wave-function as \cite{Bodwin:1994jh} \begin{eqnarray} R(0)=\frac{R_p+3R_v}{4}\end{eqnarray} where $R_p$ and $R_v$ are the normalized reduced wave-functions for pseudoscalar and vector states. The numerical results of the wave-function at the origin are given in Table \ref{tab:wave_function}.

{\renewcommand{\arraystretch}{0.8}
\begin{table*}[!ht]
\caption{$S$ and $P$ wave mass spectra of $b\bar b$ meson (in GeV)}
\begin{tabular*}{\textwidth}{@{\extracolsep{\fill}}cccccccc@{}}
\hline
State &without & with& Expt. \cite{Tanabashi:2018oca}& \cite{Ebert:2011jc}& \cite{Godfrey:2015dia}& \cite{Deng:2016ktl} & \cite{Segovia:2016xqb} \\
&correction&correction&&&&&\\
\hline
$1^1S_0$ & 9.483&9.399&9.398$\pm$0.020 & 9.398 & 9.402 & 9.390 &9.455 \\
$1^3S_1$ & 9.501&9.470&9.460$\pm$0.003 & 9.460 & 9.465& 9.460&9.502 \\
\hline
$2^1S_0$ & 9.995&9.986&9.999$\pm$0.040 & 9.990 & 9.976 & 9.990 & 9.990\\
$2^3S_1$& 10.009 &10.033&10.023$\pm$0.003 & 10.023 & 10.003 & 10.015 & 10.015\\
\hline
$3^1S_0$ & 10.290&10.315& --  & 10.329 & 10.336& 10.326 &10.330 \\
$3^3S_1$ & 10.302&10.352&10.355$\pm$0.005 & 10.355 & 10.354& 10.343&10.349 \\
\hline
$4^1S_0$ & 10.536&10.583&--  & 10.573 & 10.523 & 10.584 & -- \\
$4^3S_1$ & 10.548&10.615&10.579$\pm$0.012 & 10.586 & 10.635 & 10.597&10.607 \\
\hline
$5^1S_0$& 10.755 &10.816&--  & 10.851 & 10.869 & 10.800 &-- \\
$5^3S_1$ & 10.765&10.845&-- & 10.869 & 10.878 & 10.811&10.818 \\
\hline
$6^1S_0$& 10.953 &11.024&--  &11.061 & 11.097 & 10.997 &-- \\
$6^3S_1$& 10.962 &11.051&-- & 11.088 & 11.102 & 10.988 & 10.995\\
\hline
$1^3P_0$& 9.837 &9.837&9.859$\pm$0.004 & 9.859 & 9.847 &9.864 & 9.855\\
$1^3P_1$& 9.807 &9.852&9.892$\pm$0.003 & 9.892 & 9.876 & 9.903 &9.874 \\
$1^3P_2$& 9.823 &9.877&9.912$\pm$0.003 & 9.912 & 9.897& 9.921 &9.886 \\
$1^1P_1$& 9.852 &9.864&9.899$\pm$0.008 & 9.900& 9.882 &9.909 &9.879 \\
\hline
$2^3P_0$ & 10.247&10.258&10.232$\pm$0.004 & 10.233 & 10.226 &10.220 & 10.221\\
$2^3P_1$ & 10.206&10.279&10.255$\pm$0.002 & 10.255 & 10.246 & 10.249 &10.236\\
$2^3P_2$& 10.227 &10.317&10.268$\pm$0.002 & 10.268 & 10.261 & 10.264&10.246\\
$2^1P_1$ & 10.267&10.298&10.259$\pm$0.012 & 10.260& 10.250 & 10.254&10.240 \\
\hline
$3^3P_0$& 10.486 &10.503&-- & 10.521 & 10.522 &10.490 &10.500 \\
$3^3P_1$ & 10.433&10.529&10.513$\pm$0.007 & 10.541 & 10.538 &10.515 & 10.513\\
$3^3P_2$ & 10.460&10.580&-- & 10.550 &10.550 & 10.528& 10.521\\
$3^1P_1$& 10.512 &10.555&-- & 10.544& 10.541 &10.519 &10.516 \\
\hline
$4^3P_0$ & 10.704&10.727&--& 10.781 & 10.775 & --& --\\
$4^3P_1$& 10.644 &10.756&-- & 10.802 & 10.788 & --& --\\
$4^3P_2$& 10.672 &10.814&-- & 10.812 & 10.798 &-- & --\\
$4^1P_1$ & 10.734&10.785&-- & 10.804& 10.790 & --& --\\
\hline
$5^3P_0$ & 10.903&10.930&--& -- & 11.004 & --& --\\
$5^3P_1$& 10.839 &10.962&-- & -- & 11.014 & --& --\\
$5^3P_2$& 10.871 &11.026&-- & -- & 11.022 &-- & --\\
$5^1P_1$& 10.935 &10.994&-- & -- & 11.016 & --& --\\
\hline
\end{tabular*}
\label{Table:mass S and P}
\end{table*}
\begin{table*}[!ht]
\caption{$D$ and $F$ wave mass spectra of $b\bar b$ meson (in GeV)}
\begin{tabular*}{\textwidth}{@{\extracolsep{\fill}}cccccccc@{}}
\hline
State &without&with& Expt. \cite{Tanabashi:2018oca}& \cite{Ebert:2011jc}& \cite{Godfrey:2015dia}& \cite{Deng:2016ktl} & \cite{Segovia:2016xqb} \\
&correction&correction&&&&&\\
\hline
$1^3D_1$& 10.035&10.086&--  & 10.154 & 10.138 &10.146 &10.117 \\
$1^3D_2$& 10.073&10.123&10.163$\pm$0.014 & 10.161&10.147 &10.153 &10.122 \\
$1^3D_3$& 10.128&10.175&-- & 10.166 & 10.115 & 10.157& 10.127\\
$1^1D_2$& 10.091&10.140&--  & 10.163 & 10.148 &10.153 & 10.123\\
\hline
$2^3D_1$& 10.383&10.451&--  & 10.435 & 10.441 &10.425 & 10.414\\
$2^3D_2$& 10.431&10.497&--  & 10.443 & 10.449 & 10.432& 10.418\\
$2^3D_3$& 10.500&10.563&--  & 10.449 & 10.455 &10.436 &10.422 \\
$2^1D_2$& 10.454&10.519&--  & 10.445& 10.450 &10.432 & 10.419\\
\hline
$3^3D_1$& 10.617&10.652&--  & 10.704 & 10.698 &-- &-- \\
$3^3D_2$& 10.672&10.707&-- & 10.711 & 10.705& --&-- \\
$3^3D_3$& 10.754&10.787&--  & 10.717 & 10.711 &-- &-- \\
$3^1D_2$& 10.700&10.733&--  & 10.713 & 10.706 & --& --\\
\hline
$4^3D_1$& 10.840&10.848&--  & 10.949 & 10.928 &-- &-- \\
$4^3D_2$& 10.839&10.909&-- & 10.957 & 10.934& --&-- \\
$4^3D_3$& 10.900&11.000&--  & 10.963 & 10.939&-- &-- \\
$4^1D_2$& 10.930&10.940&--  & 10.959 & 10.935 & --& --\\
\hline
$5^3D_1$&11.034& 11.047&--  & 10.704 &--&-- &-- \\
$5^3D_2$&11.099& 11.113&-- & 10.711 &--& --&-- \\
$5^3D_3$&11.197& 11.211&--  & 10.717 & -- &-- &-- \\
$5^1D_2$&11.132& 11.145&--  & 10.713 & -- & --& --\\
\hline
$1^3F_2$& 10.223&10.294&--  & 10.343 & 10.350 & 10.338&10.315 \\
$1^3F_3$& 10.290&10.355&-- & 10.346 & 10.355& 10.340& 10.321\\
$1^3F_4$& 10.374&10.429&--  & 10.349 & 10.358 &10.340 &-- \\
$1^1F_3$& 10.309&10.372&--  & 10.374 & 10.355 & 10.339& 10.322\\
\hline
$2^3F_2$& 10.542&10.610&--  & 10.610 & 10.615 & --& --\\
$2^3F_3$& 10.619&10.692&-- & 10.614 & 10.619& --& --\\
$2^3F_4$& 10.717&10.798&--  & 10.617 & 10.622 &-- &-- \\
$2^1F_3$& 10.643&10.718&--  & 10.647 & 10.619 & --& --\\
\hline
\end{tabular*}
\label{Table:mass D and F}
\centering \raggedright{Note:- Here, in Tables \ref{Table:mass S and P} \& \ref{Table:mass D and F} the ``without correction'' row contains the masses calculated considering only the Cornell potential, and the ``with correction'' row contains the masses calculated by incorporating the relativistic correction in the framework of pNRQCD to the Cornell potential}.
\end{table*}}

\section{Regge Trajectories}
\label{sec:Regge}
Using calculated radial and orbital excited states masses of $b\bar{b}$ meson, the Regge trajectories are constructed in $(n_r,M^2)$ and $(J,M^2)$ planes with the principal quantum number related to $n_r$ via relation $n_r = n-1$ and $J$ is total angular momentum quantum number. Following equations are used
\begin{eqnarray}
 J = \alpha M^2 + \alpha_0\\
 n_r = \beta M^2 + \beta_0
\end{eqnarray}
$\alpha, \beta$ are the slopes and $\alpha_0, \beta_0$ are the intercepts.

\begin{figure}[!htb]
\centering
\includegraphics[width=8cm]{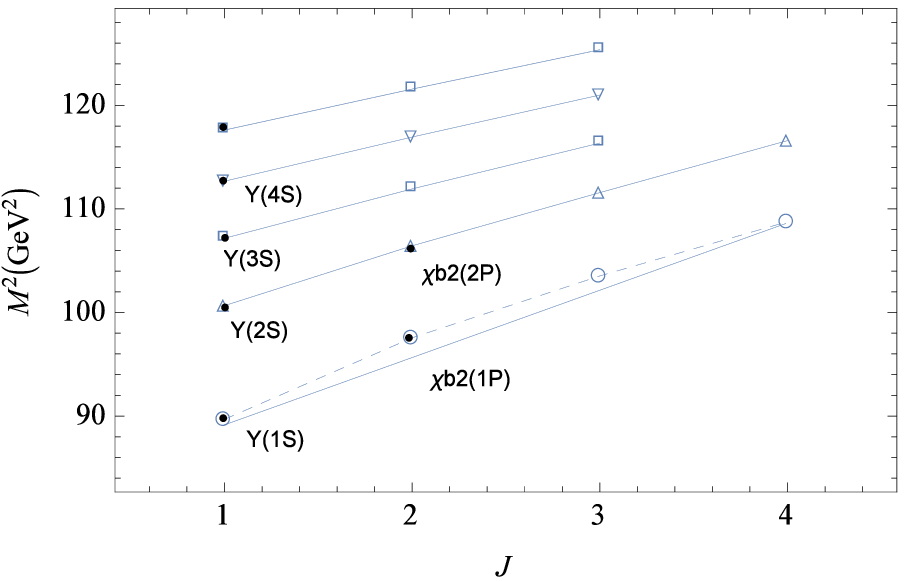}
\includegraphics[width=8cm]{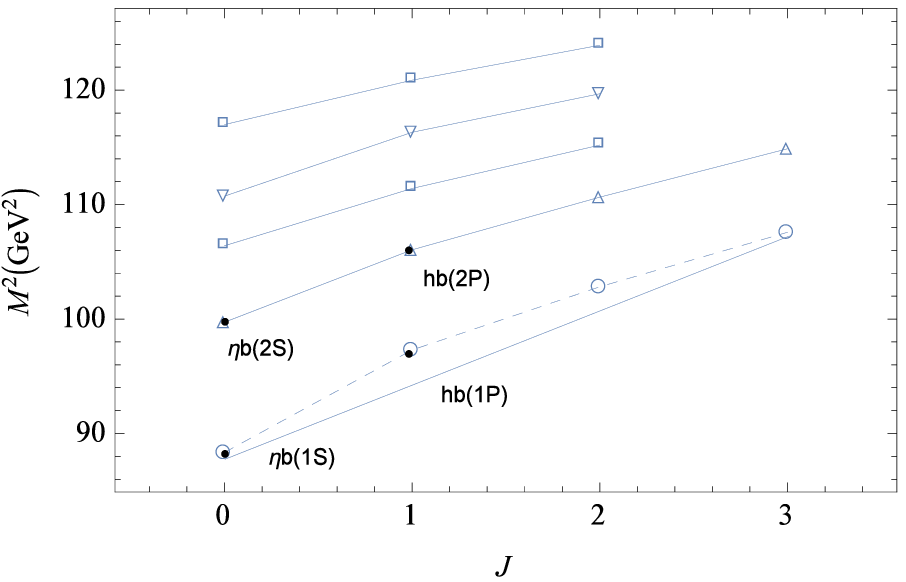}
\caption{Regge trajectory $(J,M^2)$ of $b\bar{b}$ meson with natural and unnatural parity (black dot indicates Expt. mass)}
\label{fig:1}
\end{figure}

\begin{figure}[!htb]
\includegraphics[width=8cm]{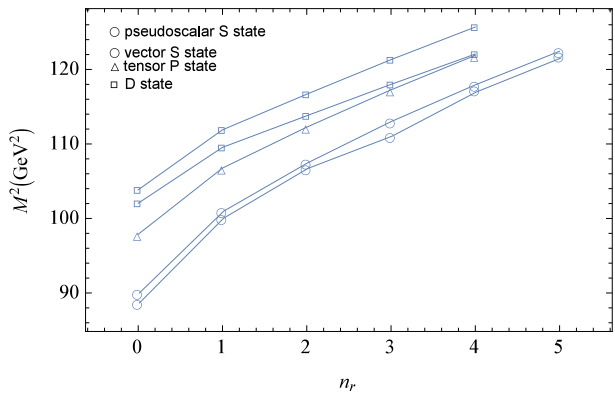}
\centering
\caption{Regge trajectory $(n_r,M^2)$ for the Pseudoscalar and vector $S$ state, excited $P$ and $D$ state masses of the $b\bar{b}$ meson}
\label{fig:2}
\end{figure}

\begin{figure}[!htb]
\centering
\includegraphics[width=8cm]{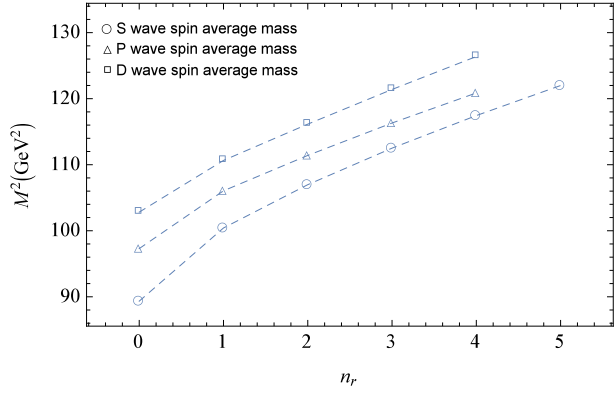}
\caption{Regge trajectory $(n_r,M^2)$ for the $S-P-D$ States center of weight mass for the $b\bar{b}$ meson}
\label{fig:3}
\end{figure}

In Figures \ref{fig:1},\ref{fig:2},\ref{fig:3} we have constructed Regge trajectories in $(n_r,M^2)$ and $(J,M^2)$ planes for $b\bar{b}$ meson. Our calculated masses are shown by solid lines and the experimental masses are shown by dots. Fitted slopes and intercepts are given in Tables \ref{Table:natural and un-natural},\ref{Table:SPD},\ref{Table:center of weight}.
Due to compactness of ground and lowest excited states, the parent trajectories of $b\bar{b}$ meson are mainly non-linear, which puts them in region where both linear confining and coulomb part of the potential play an important role. Regge trajectory can also help in the assignment of experimentally observed highly excited $b\bar{b}$ meson state to a particular $b\bar{b}$ meson state and could also help in determining their quantum numbers.

\begin{table}[!ht]
\caption{Fitted parameters of the $(J,M^2)$ Regge trajectory with natural and un-natural parity}
\begin{tabular*}{\columnwidth}{@{\extracolsep{\fill}}cccc@{}}
\hline
Parity&$b\bar b$ & $\alpha (GeV^{-2})$ & $\alpha_0$ \\
\hline
&Parent& 0.175$\pm$0.007&-15.6$\pm$0.746\\
Natural&First Daughter& 0.167$\pm$0.014&-15.5$\pm$1.515\\
&Second Daughter& 0.18$\pm$0.015&-18.4$\pm$1.725\\
&Third Daughter& 0.128&-12.855\\
\hline
&Parent&0.177$\pm$0.010&-16.705$\pm$1.059\\
Unnatural&First Daughter&0.171$\pm$0.014&-16.850$\pm$1.548\\
&Second Daughter&0.192$\pm$0.015&-19.934$\pm$1.722\\
&Third Daughter&0.137&-14.740\\
\hline
\end{tabular*}
\label{Table:natural and un-natural}
\end{table}

\begin{table}[!ht]
\caption{Fitted parameters of Regge trajectory $(n_r,M^2)$ for the Pseudoscalar and vector $S$ state, excited $P$ and $D$ state masses}
\begin{tabular*}{\columnwidth}{@{\extracolsep{\fill}}cccc@{}}
\hline
$b\bar b$& $J^{P}$ & $\beta(GeV^{-2})$ & $\beta_0$ \\
\hline
$\eta_b$&$0^{-}$&0.153$\pm$0.013&-12.913$\pm$1.433\\
$\Upsilon$&$1^{-}$&0.155$\pm$0.013&-13.332$\pm$1.474\\
$\chi_{b2}$&$2^{+}$&0.167$\pm$0.015&-15.574$\pm$1.515\\
$\Upsilon^3D_1$&$1^{-}$&0.202$\pm$0.162&-19.814$\pm$1.827\\
$\Upsilon^3D_3$&$3^{-}$&0.185$\pm$0.014&-18.378$\pm$1.656\\
\hline
\end{tabular*}
\label{Table:SPD}
\end{table}

\begin{table}[!ht]
\caption{Fitted parameters of Regge trajectory $(n_r,M^2)$ for the $S-P-D$ states center of weight mass}
\begin{tabular*}{\columnwidth}{@{\extracolsep{\fill}}ccc@{}}
\hline
$b\bar b$ & $\beta(GeV^{-2})$ & $\beta_0$ \\
\hline
S&0.154$\pm$0.013&-13.195$\pm$1.475\\
P&0.171$\pm$0.014&-15.850$\pm$1.584\\
D&0.171$\pm$0.009&-16.776$\pm$1.084\\
\hline
\end{tabular*}
\label{Table:center of weight}
\end{table}

\section{Decay Widths of $S$ and $P$ states of $b\bar{b}$ meson using NRQCD approach}
\label{sec:Decay}
Along with the masses, successful prediction of decay width is important for the validity of any model. The study of processes involving strong decays, radiative decays and leptonic decays of vector meson provides insight into quark gluon dynamics that is commendatory to what is learnt from pseudoscalar meson. Extracted model parameters and radial wave functions are employed to compute various annihilation widths. This provides insight about quark/anti-quark dynamics in bottomonia and also tests the validity of the potential.
Decay width is determined by the non-perturbative approach like NRQCD in order to test non-perturbative aspects of QCD for heavy flavor studies. It is expected that NRQCD has all corrective contribution for calculation of the decay width. Decay of heavy quarkonia is amongst the earliest application of perturbative QCD \cite{Barbieri:1975am,Appelquist:1974zd}. Short-distance factor is related to annihilation width of heavy quark anti-quark and this part is calculated in terms of running coupling constant $\alpha_s(m_Q)$, evaluated at scale of heavy-quark mass $(m_Q)$, while long-distance factor that contains all non-perturbative effects of QCD is expressed in terms of meson's non-relativistic wave function or its derivative.
In this section, we compute the $\gamma\gamma$, $e^{+}e^{-}$, light hadron and $\gamma\gamma\gamma$ decay widths considering certain relativistic orders $(\alpha_s \& \alpha_s^2)$ and various leading orders in $\nu$. The relativistic correction in computation of the decay widths might be greater in magnitude than the leading order contribution as reported in \cite{Bodwin:2002hg}. This statement on the validity of  expansion to various leading orders motivates us to investigate decays at relativistic orders and leading orders.

\subsection{$\gamma\gamma$ decay width}
The $\gamma\gamma$ decay width for $\eta_b(nS)$ states has been calculated at NNLO in $\nu$, at NLO in $\nu^2$, at $O(\alpha_s \nu^2)$ and at NLO in $\nu^4$.
NRQCD factorization expression for the decay widths of quarkonia at NLO in $\nu^4$ is given as\cite{Braaten:1995ej}
\begin{eqnarray}\label{eq:nq1}
\nonumber && \Gamma(^1 S_0 \rightarrow \gamma \gamma)  =
\frac{F_{\gamma
\gamma}(^1S_0)}{m^2_Q} \left |\left<0|\chi^{\dag}\psi|^1 S_0\right>\right|^2 \nonumber \\ && + \frac{G_{\gamma \gamma}(^1S_0)}{m^4_Q} Re \left
[\left<^1S_0|\psi^{\dag}\chi|0\right>\left<0|\chi^{\dag}\left(-\frac{i}{2}\overrightarrow{D}\right)^2\psi|^1S_0\right>\right]
\nonumber \\ && + \frac{H^1_{\gamma \gamma}(^1S_0)}{m^6_Q}\left<^1
S_0|\psi^{\dag}\left(-\frac{i}{2}\overrightarrow{D}\right)^2\chi|0\right> \left<0|\chi^{\dag}\left(-\frac{i}{2}\overrightarrow{D}\right)^2\psi|^1S_0\right> \nonumber \\ && +  \frac{H^2_{\gamma \gamma}(^1S_0)}{m^6_Q}\times
   Re\left[\left<^1S_0|\psi^{\dag}\chi|0\right>\left<0|\chi^{\dag}\left(-\frac{i}{2}\overrightarrow{D}\right)^4\psi|^1S_0\right>\right]
\end{eqnarray}
The matrix element $\left<^1S_0|{\cal{O}}(^1S_0)|^1S_0\right>$, $\left<^1S_0|{\cal{P}}_1(^1S_0)|^1S_0\right>$ and $\left<^1S_0|{\cal{Q}}^1_1(^1S_0)|^1S_0\right>$ that contribute to the decay rates of $\eta_b(nS)$ states to $\gamma \gamma$ is given as,
\begin{eqnarray}\label{eq:me1}
\left<^1S_0|{\cal{O}}(^1S_0)|^1S_0\right>&=&\left|\left<0|\chi^{\dag}\psi|^1S_0\right>\right|^2[1+
O(v^4 \Gamma)]\nonumber \\
\left<^3S_1|{\cal{O}}(^3S_1)|^3S_1\right>&=&\left|\left<0|\chi^{\dag}\sigma\psi|^3S_1\right>\right|^2[1+O(v^4 \Gamma)]\nonumber \\
\left<^1S_0|{\cal{P}}_1(^1S_0)|^1S_0\right>&=&Re\left[\left<^1S_0|\psi^{\dag}\chi|0\right> \left<0|\chi^{\dag}(-\frac{i}{2}\overrightarrow{D})^2 \psi|^1S_0\right>\right]+ O(v^4)
\end{eqnarray}
$\psi$ \& $\chi$ are Pauli spinor fields that creates heavy quark and anti-quark, $\overrightarrow{D}$ is gauge covariant spatial derivative, and $\psi^{\dag}$ \& $\chi^{\dag}$ are mixed two fermion operator corresponding to the annihilation (or creation) of $Q\bar{Q}$ pair respectively, and $\left |\left<0|\chi^{\dag}\psi|^1 S_0\right>\right|^2$, $\left
[\left<^1S_0|\psi^{\dag}\chi|0\right>\left<0|\chi^{\dag}\left(-\frac{i}{2}\overrightarrow{D}\right)^2\psi|^1S_0\right>\right]$,
$\left [\left<^1S_0|\psi^{\dag}\left(-\frac{i}{2}\overrightarrow{D}\right)^2\chi|0\right>
\left<0|\chi^{\dag}\left(-\frac{i}{2}\overrightarrow{D}\right)^2\psi|^1S_0\right> \right]$, \\ $\left[\left<^1S_0|\psi^{\dag}\chi|0\right>\left<0|\chi^{\dag}\left(-\frac{i}{2}\overrightarrow{D}\right)^4\psi|^1S_0\right>\right]$ are the operators responsible for $\gamma\gamma$ decay of $n ^1 S_0$ states, where (n=1 to 6).

The vacuum-saturation approximation expresses the four fermion matrix elements in terms of vacuum-to-quarkonium matrix elements and the heavy-quark spin-symmetry relation is used to equate the matrix elements \cite{Bodwin:1994jh}. As a result, the matrix elements are reduced to independent non-perturbative quantities  $|R_p(0)|^2$ and $|\overline{R^*_{P}}\ \overline{\bigtriangledown^2 R_{P}}|$. By expressing the decay rates in terms of matrix elements, a rigorous field theoretical definition of the non-perturbative factors in the decay rates are written as
\begin{eqnarray}\label{eq:wf1}
 \left<^1S_0|{\cal{O}}(^1S_0)|^1S_0\right>&=&\frac{3}{2 \pi}|R_{P}|^2 \nonumber \\
\left<^1S_0|{\cal{P}}_1(^1 S_0)|^1S_0\right>
&=&-\frac{3}{2 \pi}|\overline{R^*_{P}}\ \overline{\bigtriangledown^2 R_{P}}| \nonumber \\
\left<^1S_0|{\cal{Q}}^1_1(^1S_0)|^1S_0\right>&=&
-\sqrt{\frac{3}{2\pi}} \overline{\nabla^2} R_{P}
\end{eqnarray}

\begin{table*}[!htb]
\caption{Normalised reduced wave-function at the origin (in $GeV^{3/2}$)}
\renewcommand{\arraystretch}{0.7}
\begin{tabular*}{\textwidth}{@{\extracolsep{\fill}}ccc@{}}
\hline
State & $|R_p(0)|$ & $|R_v(0)|$\\
\hline
$1S$ &12.811	& 13.005\\
$2S$ &8.541 	& 8.595\\
$3S$ &6.741 	& 6.789\\
$4S$ &5.861 	& 5.869\\
$5S$ &5.310 	& 5.339\\
$6S$ &4.923 	& 4.947\\
\hline
\end{tabular*}
\label{tab:wave_function}
\end{table*}

From Equation (\ref{eq:nq1}), for calculations at leading orders in $\nu$ only the first term is considered, for calculation at leading orders at $\nu^2$ the first two terms are considered, and for calculation at leading orders at $\nu^4$ all the terms are considered.
The coefficients $F$, $G$ and $H$ in terms of various parameters are written as\cite{Brambilla:2010cs,Bodwin:1994jh,Bodwin:2002hg,Feng:2015uha,Brambilla:2018tyu,Jia:2011ah}
\begin{eqnarray}\label{eq:nq2}
F_{\gamma \gamma}(^1 S_0)&=&2\pi Q^4 \alpha^2 \bigg[ 1 + C_F {\alpha_s\over \pi}\left( \frac{\pi^2}{4}-5 \right) + \nonumber \\ &&
C_F{\alpha_s^2\over \pi^2}\bigg[C_F \left(-21-\pi^2 \left(\frac{1}{4 \epsilon} + ln\frac{\mu}{m} \right) \right) \nonumber \\ &&
 + C_A \left(-4.79 - \frac{\pi^2}{2}\left( \frac{1}{4 \epsilon} + ln\frac{\mu}{m} \right) \right) - \nonumber \\ &&
 0.565N_LT_R + 0.22N_HT_R \bigg] \bigg]
\end{eqnarray}
\begin{eqnarray}\label{eq:nq3}
G_{\gamma \gamma}(^1S_0)=
-{8\pi Q^4\alpha^2 \over 3}\bigg[1+ {C_F\alpha_s \over \pi}\bigg({5\pi^2\over 16}-{49\over 12}- \ln {\mu^2 \over 4m^2}\bigg)\bigg]
\end{eqnarray}
\begin{eqnarray}\label{eq:nq4}
H_{\gamma \gamma}(^1S_0)+H^2_{\gamma\gamma}(^1S_0)=\frac{136\pi}{45} Q^4 \alpha^2
\end{eqnarray}
$Q$ is charge of bottom quark, $\alpha$ is electromagnetic running coupling constant ($1/137$), $C_f=(N_c^2 -1)/2N_c$ is the Casimir factor for the fundamental representation and $n_f$ corresponds to the flavour of light quark. $C_F=4/3$, $C_A=3$, $N_H=1$, $T_R=1/2$, $N_L=3$ and $\mu=0.5$ with $|R_{P}(0)|$ to be the normalized wave-functions at origin.

For calculations at NNLO in $\nu$, the entire Equation (\ref{eq:nq2}) is used. For calculations at NLO in $\nu^2$ only the first two terms in the square bracket of Equation (\ref{eq:nq2}) are used, and $G_{\gamma \gamma}(^1S_0)$ is taken as $-\frac{8 \pi Q^4}{3}\alpha^2$. For NLO in $\nu^4$, in addition to first two terms in the square bracket of Equation (\ref{eq:nq2}) and $G_{\gamma \gamma}(^1S_0)= -\frac{8 \pi Q^4}{3}\alpha^2$, Equation (\ref{eq:nq4}) is also used. But, for calculation at $O(\alpha_s \nu^2)$  the first two terms in square bracket of Equation (\ref{eq:nq2}) and the entire Equation (\ref{eq:nq3}) are only used. The calculated decay widths are tabulated in Table \ref{Table:gammas}.

The decay widths for $n^3P_J (J=0,2)$ states to NLO in $\nu^2$ and NNLO in $\nu^2$ have also been calculated. $\gamma\gamma$ decay width of $n^3P_0$ and $n^3P_2$ is expressed as,
\begin{eqnarray}\label{eq:p1}
 \Gamma(\chi_{cJ} \rightarrow \gamma \gamma) &=& \frac{3 N_c Im F_{\gamma\gamma} (^3P_J)}{\pi m_Q^4},\ \ \  J = 0,2.
\end{eqnarray}
Short distance coefficients $F$'s, at NNLO in $\nu^2$ are given by \cite{Bodwin:2002hg,Sang:2015uxg}
\begin{eqnarray}\label{eq:p2}
&&F_{\gamma\gamma} (^3P_0) = 3 \pi Q^4 \alpha^2 \bigg[1+C_F {\alpha_s\over \pi}\bigg(\frac{\pi^2}{4} -\frac{7}{3}\bigg)+ \nonumber \\ && \frac{\alpha_s^2}{\pi^2}\bigg[C_F\frac{\beta_0}{4}\bigg(\frac{\pi^2}{4}-\frac{7}{3}\bigg)
\ln\frac{\mu_R^2}{m^2}\bigg]\bigg]
\end{eqnarray}
\begin{eqnarray}
F_{\gamma\gamma} (^3P_2) &=& \frac{4 \pi Q^4 \alpha^2}{5}
\bigg[1 - 4 C_f \frac{\alpha_s}{\pi} + \frac{\alpha_s^2}{\pi^2} \bigg(-2C_F\frac{\beta_0}{4}\ln\frac{\mu_R^2}{m^2}\bigg) \bigg]
\end{eqnarray}
$\beta_0 = {11\over 3}C_A - {2\over 3}(n_L+n_H)$ is
the one-loop coefficient of the QCD $\beta$-function, where
$n_H=1$, $C_A=3$ and $n_L$ signifies the number of light quark flavors. The calculated decay widths are tabulated in Table \ref{Table:gammap}.

\begin{table*}[!htb]
\caption{$\gamma\gamma$ decay widths of $\eta_b(nS)$ states(in keV)}\
\centering
\begin{tabular*}{\textwidth}{@{\extracolsep{\fill}}ccccccc}
\hline \hline
$\Gamma$ &&& State &&& \\ 
&$1^1S_0$&$2^1S_0$&$3^1S_0$&$4^1S_0$&$5^1S_0$&$6^1S_0$\\
\hline
NNLO in $\nu$&2.041&0.398&0.156&0.089&0.060&0.044\\
NLO in $\nu^2$&0.839&0.163&0.089&0.050&0.036&0.026\\
$O(\alpha_s \nu^2)$&0.858&0.166&0.091&0.051&0.037&0.026\\
NLO in $\nu^4$&0.545&0.124&0.105&0.068&0.060&0.048\\
\cite{Tanabashi:2018oca}&0.384$\pm$0.047&0.191$\pm$0.025&&&&\\
\cite{Li:2009nr}&0.527&0.263&0.172&0.105&0.121&0.050\\
\cite{Godfrey:1985xj}&0.214&0.121&0.906&0.755&&\\
\cite{Ebert:2003mu}&0.350&0.150&0.100&&&\\
\cite{Fischer:2014cfa}&0.230&0.070&0.040&&&\\
\hline \hline
\end{tabular*}
\label{Table:gammas}
\end{table*}

\begin{table*}[!htb]
\caption{$\gamma\gamma$ decay widths of $n^3P_J (J=0,2)$ states(in keV)}\
\begin{tabular*}{\textwidth}{@{\extracolsep{\fill}}ccccccccc}
\hline \hline
$\Gamma$ &&$\chi_0$&&&&$\chi_2$& \\
\hline
$\Gamma$ &1P&2P&3P&4P&1P&2P&3P&4P \\
\hline
NLO in $\nu^2$&0.069&0.022&0.012&0.008&0.014&0.004&0.002&0.002\\
NNLO in $\nu^2$&0.068&0.022&0.012&0.008&0.016&0.005&0.003&0.002\\
\cite{Li:2009nr}&0.050&0.037&0.037&&0.006&0.006&0.006&\\
\cite{Godfrey:1985xj}&0.021&0.023&&&0.005&0.006&&\\
\cite{Ebert:2003mu}&0.038&0.029&&&0.008&0.006&&\\
\hline \hline
\end{tabular*}
\label{Table:gammap}
\end{table*}

\subsection{$e^{+}e^{-}$ decay width}
The $e^{+}e^{-}$ decay width of ${{\mathit \Upsilon}{(nS)}}$ states have been calculated at NNLO in $\nu$, NLO in $\nu^2$, and at $\alpha_s^2 \nu^4$. NRQCD factorization expression for the decay widths of quarkonia at NNLO in $\nu^4$ is written as,
\begin{eqnarray} \label{eq:nq5}
&&\Gamma(^3S_1 \rightarrow
e^+e^-) = \frac{F_{ee}(^3S_1)}{m^2_Q}
\left|\left<0|\chi^{\dag}\sigma\psi|^3S_1\right>\right|^2 \cr
&&+\frac{G_{ee}(^3S_1)}{m^4_Q}
Re\left[\left<^3S_1|\psi^{\dag}\sigma\chi|0\right>\left<0|\chi^{\dag}\sigma\left(-\frac{i}{2}\overrightarrow{D}\right)^2\psi|^3S_1\right>\right]
\cr
&&+\frac{H^1_{ee}\left(^1S_0\right)}{m^6_Q}
\left<^3S_1|\psi^{\dag}\sigma\left(-\frac{i}{2}\overrightarrow{D}\right)^2\chi|0\right>
\left<0|\chi^{\dag}\sigma\left(-\frac{i}{2}\overrightarrow{D}\right)^2\psi|^3S_1\right> \nonumber \\ &&  +
 \frac{H^2_{ee}(^1 S_0)}{m^6_Q} Re \left[\left<^3S_1|\psi^{\dag}\sigma\chi|0\right>
\left<0|\chi^{\dag}\sigma\left(-\frac{i}{2}\overrightarrow{D}\right)^4\psi|^3S_1\right>\right]
\end{eqnarray}
From Equation (\ref{eq:nq5}), for calculations at leading orders in $\nu$, only the first term is considered, for calculation at leading orders at $\nu^2$ the first two terms are considered, and for calculation at leading orders at $\nu^4$ all the terms are considered.
The matrix elements that contribute to the decay rates of ${{\mathit \Upsilon}{(nS)}}  \rightarrow e^+e^-$ through the vacuum-saturation approximation gives \cite{Bodwin:1994jh}.
\begin{eqnarray}\label{eq:me2}
&&\left<^3S_1|{\cal{P}}_1(^3S_1)|^3S_1\right>=
Re\left[\left<^3S_1|\psi^{\dag}\sigma\chi|0\right> \left<0|\chi^{\dag}
\sigma\left(-\frac{i}{2}\overrightarrow{D}\right)^2\psi|^3S_1\right>\right]
+ O(v^4) \nonumber \\
&&\left<^1S_0|{\cal{Q}}^1_1(^1S_0)|^1S_0\right>=
\left<0|\chi^{\dag}\left(-\frac{i}{2}\overrightarrow{D}\right)^2\psi|^1S_0\right>\nonumber\\
&&\left<^3S_1|{\cal{Q}}^1_1(^3S_1)|^3S_1\right>=\left<0|\chi^{\dag} \sigma
\left(-\frac{i}{2}\overrightarrow{D}\right)^2\psi|^3S_1\right>
\end{eqnarray}
$\left |\left<0|\chi^{\dag}\psi|^1 S_0\right>\right|^2$, $\left[\left<^3S_1|\psi^{\dag}\sigma\chi|0\right>\left<0|\chi^{\dag}\sigma\left(-\frac{i}{2}\overrightarrow{D}\right)^2\psi|^3S_1\right>\right]$,\\ $\left[\left<0|\chi^{\dag}\sigma\left(-\frac{i}{2}\overrightarrow{D}\right)^2\psi|^3S_1\right> \right]$ and  $\left[\left<^3S_1|\psi^{\dag}\sigma\chi|0\right> \left<0|\chi^{\dag}\sigma\left(-\frac{i}{2}\overrightarrow{D}\right)^4\psi|^3S_1\right>\right]$ are the operators responsible for $e^{+}e^{-}$ decay of $n^3S_1$ states, where (n=1 to 6).

Vacuum saturation allows the matrix elements $\left<^3S_1|{\cal{O}}(^3S_1)|^3S_1\right>$, $\left<^3S_1|{\cal{P}}_1(^3S_1)|^3S_1\right>$ and $\left<^3S_1|{\cal{Q}}^1_1(^3S_1)|^3S_1\right>$  to be expressed in terms of the regularized wave-function parameters given by \cite{Bodwin:1994jh}, where $R_{V}$ is the normalised reduced wave-function for vector states.
\begin{eqnarray}\label{eq:wf2}
&&\left<^3S_1|{\cal{O}}(^3S_1)|^3S_1\right>=\frac{3}{2\pi}|R_{V}|^2 \nonumber \\
&&\left<^3S_1|{\cal{P}}_1(^3 S_1)|^3S_1\right>=-\frac{3}{2
\pi}|\overline{R^*_{V}}\ \overline{\bigtriangledown^2 R_{V}}| \nonumber \\
&&\left<^3S_1|{\cal{Q}}^1_1(^3S_1)|^3S_1\right>=-
\sqrt{\frac{3}{2\pi}} \overline{ \nabla^2} R_{V}
\end{eqnarray}
The coefficients $F$, $G$ and $H$ are written as\cite{Brambilla:2010cs,Bodwin:2002hg,Marquard:2014pea,Bodwin:1994jh}.
\begin{eqnarray}\label{eq:nq6}
&&F_{ee}(^3S_1)= \frac{2 \pi Q^2 \alpha^2}{3}  \bigg[ 1- 4 C_F
\frac{\alpha_s}{\pi}  + \nonumber \\ &&
\left[-117.46+0.82n_f+  \frac{140\pi^2}{27} ln\left(\frac{2m}{\mu_A}\right)\right]
\left(\frac{\alpha_s}{\pi}\right)^2C_F  \bigg]
\end{eqnarray}
\begin{eqnarray}\label{eq:nq7}
&&G_{ee}(^3 S_1)=-
\frac{8 \pi Q^2}{9} \alpha^2
\end{eqnarray}
\begin{eqnarray}\label{eq:nq8}
H^1_{ee}(^3S_1)+H^2_{ee}(^3S_1)=\frac{58\pi}{54} Q^2 \alpha^2
\end{eqnarray}
We have computed $\nabla^2R$ term as per \cite{Khan:1995np} $\nabla^2R = -x R \frac{M}{2}, r \rightarrow 0 $, where $x  = M - (2m_Q)$ is the binding energy of the state; $M$ is mass the system and $Q$ being charge of the bottom quark, $C_F=\frac{4}{3}$ and $\alpha=\frac{1}{137}$.\\
For calculations at NNLO in $\nu$ the entire Equation (\ref{eq:nq6}) is used. For calculations at $O(\nu^2)$ NLO only the first two terms from Equation (\ref{eq:nq6}) are used and Equation (\ref{eq:nq7}) is also used. And, for calculations at $O(\alpha_s^2 \nu^4)$ Equations (\ref{eq:nq6}), (\ref{eq:nq7}) and (\ref{eq:nq8}) are used. The calculated decay widths are tabulated in table\ref{Table:ee}.
\begin{table*}[!htb]
\caption{$e^{+}e^{-}$ decay widths of ${{\mathit \Upsilon}{(nS)}}$ states(in keV)}\
\begin{tabular*}{\textwidth}{@{\extracolsep{\fill}}ccccccc}
\hline \hline
$\Gamma$  &&& State && \\
\hline
&$1^3S_1$&$2^3S_1$&$3^3S_1$&$4^3S_1$&$5^3S_1$&$6^3S_1$\\
\hline
NNLO in $\nu$&186.561&35.591&13.855&7.880&5.299&3.905\\
NLO in $\nu^2$&1.264&0.241&0.094&0.053&0.0361&0.026\\
 $\alpha_s^2\nu^4$&1.053&0.562&0.399&0.283&0.212&0.166\\
\cite{Tanabashi:2018oca}&1.340$\pm$0.018&0.612$\pm$0.011&0.443$\pm$0.008&0.272$\pm$0.029&&\\
\cite{Shah:2012js}&1.20&0.52&0.33&0.24&0.19&0.16\\
\cite{Radford:2009qi}&1.33&0.62&0.48&0.40&&\\
\cite{Patel:2008na}&1.61&0.87&0.66&0.53&0.44&0.39\\
\cite{Ebert:2003rh}&1.3&0.5&&&&\\
\cite{Gonzalez:2003gx}&0.98&0.41&0.27&0.20&0.16&0.12\\
\hline \hline
\end{tabular*}
\label{Table:ee}
\end{table*}

\subsection{Light hadron decay width}
The Light hadron decay width through NLO and NNLO in $\nu^2$ is calculated. The methodology for calculation is given as \cite{Bodwin:1994jh,Braaten:1995ej}, $R_p$ is the normalised reduced wave-function for pseudoscalar states.
\begin{eqnarray}
\Gamma(^1S_0 \rightarrow LH) = \frac{N_c \mathrm{Im}~f_1(^1S_0)}{\pi m_Q^2} |\bar{{R_p}}|^2 + \frac{N_c \mathrm{Im}~g_1(^1S_0)}{\pi m_Q^4} Re(\bar{R_p}^* \bar{\nabla^2 R_p})
\end{eqnarray}
The coefficients $F$ and $G$ for decay width calculation at NNLO at $\nu^2$ are written as \cite{Bodwin:1994jh,Feng:2017hlu},
\begin{eqnarray}\label{eq:nq9}
\mathrm{Im}~f_1(^1S_0)&=&\frac{\pi C_F \alpha_s^2}{N_c} \bigg[ 1+\frac{\alpha_s}{\pi}\bigg(\frac{\beta_0}{2}ln\frac{\mu_R^2}{4m^2}+\bigg(\frac{\pi^2}{4}-5 \bigg)C_F\nonumber \\
&& + \bigg(\frac{199}{18}-\frac{13\pi^2}{24} \bigg)C_A -\frac{8}{9}n_L - \frac{2n_H}{3}  \bigg)ln2 + \nonumber \\
&&
\frac{\alpha_s^2}{\pi^2} \bigg(-69.5 + \frac{3\beta_0^2}{16} ln^2\bigg(\frac{\mu_R^2}{4m^2}\bigg) + \bigg(\frac{\beta_1}{8}+ \frac{3}{4}\beta_0 16.20 \bigg)ln\frac{\mu_R^2}{4m^2} - \nonumber \\
&&\pi^2\bigg(C_F^2 +\frac{C_A C_F}{2}  \bigg)ln\frac{\mu_{\Lambda}^2}{m^2} \bigg) \bigg]
\end{eqnarray}
\begin{eqnarray}\label{eq:nq10}
\mathrm{Im}~g_1(^1S_0)&=&-\frac{4\pi C_F \alpha_s^2}{3N_c}\bigg[ 1+\frac{\alpha_s}{\pi}\bigg(\frac{\beta_0}{2}ln\frac{\mu_R^2}{4m^2}-C_F ln\frac{\mu_{\Lambda}^2}{m^2} \bigg)-\nonumber \\
&&\bigg(\frac{49}{12}-\frac{5\pi^2}{16} -2ln2 \bigg)C_F + \nonumber \\
&& \bigg(\frac{479}{36}-\frac{11\pi^2}{16}\bigg)C_A - \frac{41}{36}n_L- \frac{2n_H}{3}ln2 \bigg]
\end{eqnarray}
Here, $\beta_0=\frac{11}{3}C_A-\frac{4}{3}T_Fn_f$, $T_F=1/2$, $n_f=n_L+n_H$ signifies number of active flavour quark, $n_L=3$, $n_H=1$, $C_F=\frac{N_C^2-1}{2N_C}$, $C_A=N_C=3$, $\mu_R$ is the renormalisation scale,  $\beta_1=\frac{34}{3}C_A^2-\frac{20}{3}C_AT_Fn_f-4C_AT_Fn_f$ is the two-loop coefficient of the QCD $\beta$ function.

For decay width calculation at NLO in $\nu^2$ only the first two terms in the square bracket from Equation (\ref{eq:nq9}) is considered and the entire Equation (\ref{eq:nq10}) is considered. The results of the decay width is tabulated in table \ref{Table:lh}
\begin{table*}[!htb]
\caption{Light hadrons(LH) decay width of $\eta_b(nS)$ states(in MeV)}\
\begin{tabular*}{\textwidth}{@{\extracolsep{\fill}}ccccccc}
\hline
$\Gamma$&$1^1S_0$&$2^1S_0$&$3^1S_0$&$4^1S_0$&$5^1S_0$&$6^1S_0$\\
\hline
NLO in $\nu^2$ &24.755&9.900&5.810&4.214&3.272&2.706\\
NNLO in $\nu^2$ &59.056&23.664&13.886&10.071&7.819&6.465\\
\cite{Li:2012rn}&$9.76{}^{+0.58}_{-0.54}$&&&&&\\
\cite{Mizuk:2012pb}&$10.8{}^{+4.0+4.5}_{-3.7-2.0}$&&&&&\\
\hline
\end{tabular*}
\label{Table:lh}
\end{table*}

\subsection{$\gamma\gamma\gamma$ decay width}
$\gamma\gamma\gamma$ decay width of ${{\mathit \Upsilon}{(nS)}}$ states given in \cite{Mackenzie:1981sf,Bodwin:1994jh} through NLO in $\nu^2$ is also calculated and is represented by,
\begin{equation}\label{eq:nq11}
\Gamma(^3S_1 \rightarrow \gamma\gamma\gamma) = {8 (\pi^2 - 9) Q^6 \alpha^3 \over 9 \pi m_Q^2}
\Bigg[ 1 \;-\; 9.46(2) C_F {\alpha_s \over \pi} \Bigg]R_v \;
\end{equation}
Here, $\alpha = e^2/4 \pi$ and $R_v$ is the normalised reduced wavefunction for vector states.
\begin{table*}[!htb]
\caption{$\gamma\gamma\gamma$ decay width of ${{\mathit \Upsilon}{(nS)}}$ states(in eV)}
\begin{tabular*}{\textwidth}{@{\extracolsep{\fill}}ccccccc}
\hline \hline
$\Gamma$ &&&& State && \\
\hline
&$1^3S_1$&$2^3S_1$&$3^3S_1$&$4^3S_1$&$5^3S_1$&$6^3S_1$ \\
\hline
NLO in $\nu^2$&0.016&0.003&0.001&&\\
\hline \hline
\end{tabular*}
\label{Table:ggg}
\end{table*}

\section{Mixed bottomonium states}
\label{sec:Mixing}
Many hadronic states that are observed but their structure is uncertain may be admixture of nearby iso-parity states. Mass of a mixed state ($M_{nL}$) is expressed in terms of two mixing states ($nl$ and $n'l'$) as
\begin{equation}
M_{nL} = \mid a^2\mid M_{nl} + ( 1-\mid a^2\mid ) M_{n'l'}
\end{equation}
Where,$\mid a^2\mid$ $=$ $\cos^2 \theta$ and $\theta$ is mixing angle. With the help of this equation, we can obtain mixed state configuration and mixing angle \cite{Shah:2012js,Radford:2009qi}.
Computed masses and their leptonic decay width of the $S-D$ wave admixture states namely $\Upsilon(10.860)$ (mass in GeV) and $\Upsilon(11.020)$ are presented in Table \ref{Table:9}. Admixture of nearby $P$-waves for predictions of $X$(10.610) is presented in Table \ref{Table:10} along with other theoretical calculations as well as available experimentally masses and decay widths \cite{Shah:2014yma,Patel:2016otd,Yazarloo:2016zer,Bhavsar:2018umj}. For $D$-wave states their wave function at origin is defined here as in \cite{Novikov,Badalian:2009bu},
\begin{equation}
\label{eq.14} R_D(0)= \frac{5R_D''(0)}{2\sqrt{2}\omega_b^2},
\end{equation}
Here, $R_D''(0)$ is the second order derivative of the wave function at origin for $D$ state and $\omega_b$ is a constant having value 5.11 GeV \cite{Badalian:2009bu}.
Mixed $P$ wave states can be expressed as,
\begin{equation}
|\alpha\rangle=\sqrt{\frac{2}{3}}|^3P_1\rangle+\sqrt{\frac{1}{3}}|^1P_1\rangle
\end{equation}
\begin{equation}
|\beta\rangle=-\sqrt{\frac{1}{3}}|^3P_1\rangle+\sqrt{\frac{2}{3}}|^1P_1\rangle
\end{equation}
Where, $|\alpha\rangle$, $|\beta\rangle$ are states having same parity.
We can write the masses of these states in terms of the predicted masses of pure $P$ wave states ($^3P_1$ and $^1P_1$) as \cite{Shah:2014yma,Patel:2016otd,Yazarloo:2016zer},

\begin{table*}[!htb]
\caption{Mixing angle and leptonic decay width of S-D wave admixture states}
\begin{tabular*}{\textwidth}{@{\extracolsep{\fill}}ccccccc@{}}
\hline
Expt. state & J$^P$ & Mixed state & \multicolumn {2} {c} {Masses mixed state(GeV)}& \multicolumn {2} {c} {$\Gamma_{e^{+}e^{-}}$ mixed state(keV)}\\
\cline{4-7} &&& {Present} & {Expt. \cite{Tanabashi:2018oca}}&{Present}&{Expt. \cite{Tanabashi:2018oca}}\\
\hline
$\Upsilon$(10.860)& $1^-$ & $5^3S_1$ and $5^3D_1$ &10.907&10.889$\pm$0.004&0.131&0.31$\pm$0.07\\

$\Upsilon$(11.020)& $1^-$ & $6^3S_1$ and $5^3D_1$ &11.042&10.992$\pm$0.010&0.100&0.132$\pm$0.024\\
\hline
\end{tabular*}
\label{Table:9}
\end{table*}

\begin{table}[!ht]
\caption{Masses of mixed P-wave positive parity states}
\begin{tabular*}{\columnwidth}{@{\extracolsep{\fill}}cccc@{}}
\hline
Expt. State & Mixed State Configuration & Present & Expt. \cite{Tanabashi:2018oca}\\
\hline
X(10.610)& $4^3P_1$ and $3^1P_1$ & 10.622 & 10.609$\pm$0.006\\
\hline
\end{tabular*}
\label{Table:10}
\end{table}

\section{Electromagnetic transition widths}
\label{sec:EM}
Electromagnetic transitions can be determined broadly in terms of electric and magnetic multipole expansions and their study can help in understanding the non-perturbative regime of QCD. Here the electromagnetic transitions are calculated in the framework of pNRQCD.
We consider leading order terms i.e. electric ($E1$) and magnetic ($M1$) dipoles with selection rules $\Delta L = \pm 1$ and $\Delta S = 0$ for $E1$ transitions while $\Delta L = 0$ and $\Delta S = \pm 1$ for $M1$ transitions.
We employ numerical wave function for computing electromagnetic transition widths for mesonic states in order to test parameters used in the present work. For $M1$ transition, we restrict our calculations for transitions among $S$ waves only.
In non-relativistic limit, the radiative $E1$ and $M1$ transition widths are given by \cite{Brambilla:2010cs,Radford:2009qi,Eichten:1974af,Eichten:1978tg,Pandya:2014qma}

\begin{eqnarray}
 \Gamma(n^{2S+1}L_{iJ_i} \to n^{2S+1}L_{fJ_f} + \gamma) &=& \frac{4 \alpha_e \langle e_Q\rangle ^2\omega^3}{3} (2 J_f + 1) S_{if}^{E1} |M_{if}^{E1}|^2
 \end{eqnarray}
\begin{eqnarray}
\Gamma(n^3S_1 \to {n'}^{1}S_0+ \gamma) = \frac{\alpha_e \mu^2 \omega^3}{3} (2 J_f + 1) S_{if}^{M1} |M_{if}^{M1}|^2
\end{eqnarray}
where, mean charge content $\langle e_Q \rangle$ of the $Q\bar{Q}$ system, magnetic dipole moment $\mu$ and photon energy $\omega$ are given by
\begin{equation}
\langle e_Q \rangle = \left |\frac{m_{\bar{Q}} e_Q - e_{\bar{Q}} m_Q}{m_Q + m_{\bar{Q}}}\right |
\end{equation}
\begin{equation}
\mu = \frac{e_Q}{m_Q} - \frac{e_{\bar{Q}}}{m_{\bar{Q}}}
\end{equation}
and
\begin{equation}
\omega = \frac{M_i^2 - M_f^2}{2 M_i}
\end{equation}
respectively. Also, the symmetric statistical factors are given by
\begin{equation}
S_{if}^{E1} = {\rm max}(L_i, L_f)
\left\{ \begin{array}{ccc} J_i & 1 & J_f \\ L_f & S & L_i \end{array} \right\}^2\\
\end{equation}
and
\begin{equation}
S_{if}^{M1} = 6 (2 S_i + 1) (2 S_f + 1)
\left\{ \begin{array}{ccc} J_i & 1 & J_f \\ S_f & \ell & S_i \end{array} \right\}^2 \left\{ \begin{array}{ccc} 1 & \frac{1}{2} & \frac{1}{2} \\ \frac{1}{2} & S_f & S_i \end{array} \right\}^2.
\end{equation}
The matrix element $|M_{if}|$ for $E1$ and $M1$ transitions can be written as
\begin{equation}
\left |M_{if}^{E1}\right | = \frac{3}{\omega} \left\langle f \left | \frac{\omega r}{2} j_0 \left(\frac{\omega r}{2}\right) - j_1 \left(\frac{\omega r}{2}\right) \right | i \right\rangle
\end{equation}
and
\begin{equation}
\left |M_{if}^{M1}\right | = \left\langle f\left | j_0 \left(\frac{\omega r}{2}\right) \right | i \right\rangle
\end{equation}
The electromagnetic transition widths are listed in Table \ref{Table:11} and \ref{Table:12} in comparison with experimental results as well as with other theoretical predictions.
\begin{table*}[!htb]
\caption{$E1$ transition width of Bottomonia (in keV)}
\begin{tabular*}{\textwidth}{@{\extracolsep{\fill}}ccccccc@{}}
\hline
Transition & Present & Expt. \cite{Tanabashi:2018oca} & \cite{Radford:2009qi} & \cite{Ebert:2002pp} & \cite{Li:2009nr} & \cite{Deng:2016ktl}\\
\hline
$2^3S_1 \rightarrow 1^3P_0$&1.300 &1.22$\pm$0.11 &1.15 &1.65 &1.67 &1.09 \\
$2^3S_1 \rightarrow 1^3P_1$&2.232 &2.21$\pm$0.19 &1.87 &2.57 &2.54 &2.17 \\
$2^3S_1 \rightarrow 1^3P_2$&2.274 &2.29$\pm$0.20 &1.88 &2.53 &2.62 &2.62 \\
$2^1S_0 \rightarrow 1^1P_1$&8.045 &-- &4.17 &3.25 &6.10 &3.41 \\
\hline
$3^3S_1 \rightarrow 2^3P_0$&1.102 &1.20$\pm$0.12 &1.67 &1.65 &1.83 &1.21 \\
$3^3S_1 \rightarrow 2^3P_1$&1.345 &2.56$\pm$0.26 &2.74 &2.65 &2.96 &2.61 \\
$3^3S_1 \rightarrow 2^3P_2$&0.706 &2.66$\pm$0.27 &2.80 &2.89 &3.23 &3.16 \\
$3^1S_0 \rightarrow 2^1P_1$&0.093 &-- &-- &3.07 &11.0 &4.25 \\
\hline
$1^3P_0 \rightarrow 1^3S_1$&29.068 &-- &22.1 &42.7 &26.6 &27.5 \\
$1^3P_1 \rightarrow 1^3S_1$&42.551 &-- &27.3 &37.1 &33.6 &31.9 \\
$1^3P_2 \rightarrow 1^3S_1$&39.266 &-- &31.2 &29.5 &38.2 &31.8 \\
$1^1P_1 \rightarrow 1^1S_0$&74.633 &-- &37.9 &54.4 &55.8 &35.8 \\
\hline
$2^3P_0 \rightarrow 2^3S_1$&16.842 &-- &9.9 &11.7 &11.7 &14.4 \\
$2^3P_1 \rightarrow 2^3S_1$&21.844 &19.4$\pm$5.0 &13.7&15.9 &15.9 &15.3 \\
$2^3P_2 \rightarrow 2^3S_1$&15.665 &15.1$\pm$5.6 &16.8 &18.8 &18.8 &15.5 \\
$2^1P_1 \rightarrow 2^1S_0$&23.108 &-- &-- &23.6 &24.7 &16.2 \\
\hline
$2^3P_0 \rightarrow 1^3S_1$&8.715 &-- &6.69 &7.36 &11.4 &5.4 \\
$2^3P_1 \rightarrow 1^3S_1$&9.393 &8.9$\pm$2.2 &7.31 &8.01 &12.4 &10.8 \\
$2^3P_2 \rightarrow 1^3S_1$&10.704 &9.8$\pm$2.3 &7.74 &8.41 &13.0 &12.5 \\
$2^1P_1 \rightarrow 1^1S_0$&12.664 &-- &-- &9.9 &15.9 &16.1 \\
\hline
$1^3D_1 \rightarrow 1^3P_0$&3.120 &-- &-- &24.2 &23.6 &19.8 \\
$1^3D_1 \rightarrow 1^3P_1$&2.624 &-- &-- &12.9 &12.3 &13.3 \\
$1^3D_1 \rightarrow 1^3P_2$&1.908 &-- &-- &0.67 &0.65 &1.02 \\
$1^3D_2 \rightarrow 1^3P_1$&3.940 &-- &19.3 &24.8 &23.8 &21.8 \\
$1^3D_2 \rightarrow 1^3P_2$&3.017 &-- &5.07 &6.45 &6.29 &7.23 \\
$1^3D_3 \rightarrow 1^3P_2$&5.099 &-- &62.7 &252 &284 &350 \\
$1^1D_2 \rightarrow 1^1P_1$&5.523 &-- &-- &335 &575 &362 \\
\hline
\end{tabular*}
\label{Table:11}
\end{table*}

\begin{table*}[!htb]
\caption{$M1$ transition width of Bottomonia (in keV)}
\begin{tabular*}{\textwidth}{@{\extracolsep{\fill}}cccccc@{}}
\hline
Transition & Present & Expt. \cite{Tanabashi:2018oca} & \cite{Radford:2009qi} & \cite{Ebert:2002pp} & \cite{Deng:2016ktl} \\
\hline
$1^3S_1 \rightarrow 1^1S_0$&2.527 &-- &4.0 &5.8 &10 \\
$2^3S_1 \rightarrow 2^1S_0$&0.306 &-- &0.05 &1.40 &0.59 \\
$2^3S_1 \rightarrow 1^1S_0$&11.954 &12.5$\pm$4.9 &0.0 &6.4 &66 \\
$3^3S_1 \rightarrow 3^1S_0$&0.024 &-- &-- &0.8 &3.9 \\
$3^3S_1 \rightarrow 2^1S_0$&0.318 &$\leq$14 &-- &1.5 &11 \\
$3^3S_1 \rightarrow 1^1S_0$&8.452 &10$\pm$2 &-- &10.5 &71 \\
\hline
\end{tabular*}
\label{Table:12}
\end{table*}

\section{Results and Discussion}
\label{sec:Result}
We have computed the mass spectra of $1S-6S$, $1P-5P$, $1D-5D$ and $1F-2F$ states of $b\bar{b}$ meson by solving the Schr\"{o}dinger equation numerically taking into consideration the coulomb plus linear (Cornell) potential along with relativistic correction to mass in the framework of pNRQCD which is usually classified in powers of the inverse of heavy quark mass or velocity.
Confinement strength is fixed by the experimental ground state masses. In order to obtain mass splitting between degenerate mesonic states, we considered spin dependent part of the confined one gluon exchange potential along with scaling of the potential strength parameter to account for increase in average kinetic energies of the constituent quark and antiquark. The calculated masses are compared with PDG data and also with other models.

The computed masses of $S$, $P$, $D$ and $F$ states are tabulated in Tables \ref{Table:mass S and P} and \ref{Table:mass D and F}, along with latest experimental data and other theoretical approaches. $\chi^2$/\textit{d.o.f} or the goodness of fit test is an statistical test helping the analyst to get an idea about how close the observed values are to expected or true values. If the computed  $\chi^2$/\textit{d.o.f} value is large, then the observed and expected values are not close and the model is a poor fit to the data.  It is calculated as per \cite{Wuensch2011}, $\chi^2/\textit{d.o.f}= \sum \frac{(observed-expected)^2}{expected}$. In present article the calculated $\chi^2$/\textit{d.o.f} value for the parameters is $0.764$ implying that the model parameters have been appropriately chosen. To calculate this $\chi^2/\textit{d.o.f}$ value, the expected values of the potential parameters are taken as per column two and the observed values as per column one of Table \ref{Table:parameters}.

It can be observed that calculated masses are in good agreement with PDG data. Mass difference between the $S$ wave $b\bar{b}$ meson $1^1S_0$ - $1^3S_1$ is 71 MeV and $2^1S_0$ - $2^3S_1$ is 47 MeV while that from the experimental data is 62 MeV and 24 MeV respectively. Calculated mass of $3^3S_1$ and $4^3S_1$ shows 3 MeV and 36 MeV deviation respectively when compared with the experimental values. Calculated masses are consistent with the relativistic approach. The inclusion of $1/m$ dependent relativistic correction term $V_p(r)$ significantly increases the mass difference between vector and pseudoscalar $S$-wave states. While comparing our $P$-state masses with PDG data, we observe that our $1P$ sate varies in the range of 0.2 to 0.4\%, $2P$ state varies in the range of 0.25 to 0.46\%. Only one $3P$ and $1D$ state is reported in PDG i.e. $3^3P_1$ and $1^3D_2$ our calculated value differs from it only by 0.2\% and 0.4\% respectively. The $V_p(r)$ term in potential increases the mass splitting of triplet states and shifts the singlet state masses upwards making them comparable with available experimental data as well as predictions of other models. We compare the calculated masses of all other states with other theoretical approaches as they are yet to be observed in experiments and find them to be relatively varying in the range of 0.5 to 1\%.

In order to assign experimentally observed highly excited $b\bar{b}$ meson state to a particular $b\bar{b}$ meson state, we construct $(n_r,M^2)$ and $(J,M^2)$ Regge trajectories. The parent nuclide trajectory i.e. the ground state and first few excited states of $b\bar{b}$ meson are non-linear due to dominance of confining potential in the inter-quark potential and as a result, the trajectory is non-linear, while the daughter trajectories are linear because coulombic potential is less dominant. Slopes and intercepts of both linear and non-linear Regge trajectories are determined and tabulated in Tables \ref{Table:natural and un-natural},\ref{Table:SPD},\ref{Table:center of weight} respectively.
We compare slopes of linear Regge trajectory of $b\bar{b}$ meson obtained in this paper to about 0.175 $GeV^{-2}$ with the slope of $c\bar{c}$ meson from our previous work \cite{Chaturvedi:2018xrg} which is 0.27 $GeV^{-2}$ and would comment that the slope of the meson Regge trajectory is mainly determined by and depends upon the mass of the quark.

Decays provide deeper insight about the exact nature of the inter-quark forces. Using the potential parameters and reduced normalized wave functions, we compute various decay properties of $b\bar{b}$ meson. The $\gamma\gamma$ decay widths of $\eta_b (nS)$ are calculated at NNLO in $\nu$, at NLO in $\nu^2$, at $O(\alpha_s \nu^2)$ and at NLO in $\nu^4$. The results are tabulated in Table \ref{Table:gammas} in comparison with experimental and other theoretical decay widths. It can be observed that out of all the four approaches, the decay width calculated at NLO in $\nu^4$ is more desirable and nearest to the experimental determined results, this is because of the inclusion of the coefficients $H_{\gamma \gamma}(^1S_0)+H^2_{\gamma\gamma}(^1S_0)$ for calculation of the decay at NLO in $\nu^4$. here, the calculated value for these coefficients is found to be $6.24$. The calculated decay width at NLO in $\nu^4$ vary only by 0.161 and 0.067 keV respectively as compared to PDG \cite{Tanabashi:2018oca}. The calculation of $\gamma\gamma$ decay width for $\eta_b (nS)$ states is consistent at NLO in $\nu^4$ in contrast with relativistic corrections to various orders of $\alpha_s$ because the former consists of additional coefficients that are noticeable. The $\gamma\gamma$ decay widths of $n^3P_J (J=0,2)$ states are also calculated at NLO in $\nu^2$ and NNLO in $\nu^2$, the results are tabulated in Table\ref{Table:gammap}, and compared with experimental other theoretical decay width. It can be observed that, there is no difference between the decay width calculated by both the approaches and the results obtained are comparable with that of the other theoretical approaches, the width decreases with radial excitations. Also, it can be seen that the perturbative corrections of relative order $\alpha_s^2$ or NNLO in $\nu^2$ do not provide any significant change in the calculated decay width in absence of term available for calculating decay at $\nu^4$.

The $e^{+}e^{-}$ decay widths of $\Upsilon (nS)$ are calculated at NNLO in $\nu$, NLO in $\nu^2$, and at $\alpha_s^2 \nu^4$. The results are presented in Table \ref{Table:ee} along with experimental and other theoretical decay widths. The decay width calculated at NNLO in $\nu$ is significantly larger than the decay width calculated by other two approaches and is not comparable i.e. approximately two orders greater than experimental data. The decay widths calculated at $\alpha_s^2\nu^4$ are close to the experimentally determined decay width with exception for $\Upsilon (S)$. We consider calculation at $\alpha_s^2\nu^4$ appropriate as it contains additional parameters $H^1_{ee}(^3S_1)+H^2_{ee}(^3S_1)$, the computed value for these parameters is $19.97$. Also, the calculation here at $\alpha_s^2\nu^4$ contain relativistic correction terms to the order of $\alpha^2.$

The Light hadrons(LH) decay width of $\eta_b(nS)$ states is calculated at NLO and NNLO in $\nu^2$, the results are tabulated in Table\ref{Table:lh}. The decay width calculated at NLO in $\nu^2$ is suppressed as compared with the decay width calculated at NNLO in $\nu^2$, and is also nearby the decay width calculated by \cite{Li:2012rn} and \cite{Mizuk:2012pb}. The $\gamma\gamma\gamma$ decay width of ${{\mathit \Upsilon}{(nS)}}$ states calculated at NLO in $\nu^2$ is tabulated in Table \ref{Table:ggg}. Also ratio of $\frac{\Gamma_{ee}(nS)}{\Gamma_{ee}(1S)}$ in table \ref{Table:13} is consistent with experimental data \cite{Tanabashi:2018oca}. It can be commented, that potential employed here is successful in determining mass spectra and decays of the $b\bar{b}$ meson.
\begin{table}[!ht]
\caption{The ratios of $\frac{\Gamma_{ee}(nS)}{\Gamma_{ee}(1S)}$ for bottomonium states}
\begin{tabular*}{\columnwidth}{@{\extracolsep{\fill}}cccc@{}}
\hline
$\frac{\Gamma_{e^{+}e^{-}}\Upsilon(nS)}{\Gamma_{e^{+}e^{-}}\Upsilon(1S)}$&Present&Expt. \cite{Tanabashi:2018oca}& \cite{Fu:2010gd}\\
\hline
$\frac{\Gamma_{e^{+}e^{-}}\Upsilon(2S)}{\Gamma_{e^{+}e^{-}}\Upsilon(1S)}$&0.48&0.46&0.50\\
$\frac{\Gamma_{e^{+}e^{-}}\Upsilon(3S)}{\Gamma_{e^{+}e^{-}}\Upsilon(1S)}$&0.33&0.33&0.36\\
$\frac{\Gamma_{e^{+}e^{-}}\Upsilon(4S)}{\Gamma_{e^{+}e^{-}}\Upsilon(1S)}$&0.23&0.20&0.29\\
$\frac{\Gamma_{e^{+}e^{-}}\Upsilon(5S)}{\Gamma_{e^{+}e^{-}}\Upsilon(1S)}$&0.17&--&0.24\\
\hline
\end{tabular*}
\label{Table:13}
\end{table}
A detailed comparison of calculated $b\bar{b}$ meson mass spectra with experimental data is carried out for states above and below open flavour production threshold. For states below threshold, the masses are well reproduced in present study while for those above threshold, computed masses show discrepancy of few MeV. It is also shown that these states fit well into the corresponding Regge trajectory. In addition to this, other states having complex properties like $X$ and $\Upsilon$ bottomonium-like states are discussed with the possibility of mixing of $n^3S_1$ and $n^3D_1$ i.e. negative parity states and $n^3P_1$ and $n^1P_1$ i.e. positive parity states and are listed in Tables \ref{Table:9} \& \ref{Table:10}. The $e^{+}e^{-}$ decay with of negative parity admixtures is also listed in Table \ref{Table:9}. $\Upsilon(10.860)$(GeV) and $\Upsilon(11.042)$ states as admixtures of $5^3S_1$ and $5^3D_1$ states and $6^3S_1$ and $5^3D_1$ states respectively are calculated and the corresponding masses are 10.907 GeV and 11.042 GeV, respectively. The experimentally determined masses are 10.889$\pm$0.004 GeV and 10.992$\pm$0.010 GeV, our masses differ only by 18 MeV and 50 MeV respectively. The decay width of admixture state found by us is 0.131 KeV and 0.100 keV while the value of experimentally determined decay width is 0.31$\pm$0.07 KeV and 0.132$\pm$0.024 KeV respectively, thus the determined decay width is nearly equal to experimental value and we conclude that $\Upsilon(10.860)$ is mixed state of $5^3S_1$ and $5^3D_1$ and cannot be assigned as a pure $5^3S_1$ bottomonium state, also $\Upsilon(11.042)$ is mixed state of $6^3S_1$ and $5^3D_1$ and cannot be assigned as a pure $6^3S_1$ bottomonium state. In addition to these negative parity states, a positive parity state $X$(10610) has also been studied and the theoretical mass calculated by us is 10.622 GeV and experimentally determined mass is 10.609$\pm$0.006 GeV, the difference is only of 13 MeV in between two masses.

We have also calculated $E1$ and $M1$ transition for bottomonium and results for the same has been listed in Tables \ref{Table:11} \& \ref{Table:12}. The results obtained by us are compared with experimental data as well as other theoretical approaches like relativistic potential model, quark model and non-relativistic screened potential model. It is seen that for experimentally observed transition values, our calculated transition widths show very less deviation and are in fair agreement with experimental and theoretical values. Also for transition widths which are not observed experimentally our calculated values are comparable with values obtained by other theoretical approaches.
\section{Conclusion}
\label{sec:Conclusion}
We have computed mass spectra of the $b\bar{b}$ meson by taking into consideration a relativistic correction in the framework of pNRQCD to the Cornell potential. We would like to draw an important conclusion from our study of the mass spectra, as can be observed in Tables \ref{Table:mass S and P} \& \ref{Table:mass D and F} that the computed masses by considering the relativistic correction are slightly less than the masses calculated by considering only the Cornell potential for $n=1,2$ ($S$ states only). However, for states $n \geq 3 (S-states)$ and all the $P,D \& F$ states, after incorporating the relativistic correction, the masses turn out to be on higher side. Also, after considering the coupled channel effect in the screening potential, the potential at large quark anti-quark separation is modified and the calculated masses for states $n \geq 3$ drop off by a few eV as a result. However, after incorporating the relativistic correction term, the calculated masses for higher bottomonium states increase due to increase in the numerical value of $\frac{1}{m_b}V^{(1)}(r)$ term in Eq.\ref{eq:potential} for the higher states as a result of increasing quark anti-quark separation. $\gamma\gamma$, $e^+e^-$, LH and $\gamma\gamma\gamma$ decay widths are calculated using the same set of chosen parameters and the obtained reduced wave function.
We have also tried to emphasize on bottomonium like $X$(10610), $\Upsilon$(10860) and $\Upsilon$(11020) experimentally observed states as admixtures of $n^3S_1$ and $n^3D_1$ and $n^3P_1$ and $n^1P_1$ states respectively and have calculated their $e^+e^-$ decay width and electromagnetic transition widths as well. Decay properties throw light on identification of these states as admixture states. Also the Regge trajectories are constructed which are helpful for the association of higher excited states obtained into the family of $b\bar{b}$ meson. Overall results from the present study are satisfactory when compared with the latest experimental results both for the mass spectra, decays and electromagnetic transitions point of view.

\bibliography{reference}

\end{document}